\documentstyle[11pt]{article}

\input{tcilatex}
\begin{document}

\title{{\bf \ {\large Comment on Ricci Collineations for type B warped space-times} 
}}
\author{M Tsamparlis\thanks{%
Department of Physics, Section of Astronomy-Astrophysics-Mechanics,
University of Athens, ZOGRAFOS 15783, ATHENS, GREECE.
E-mails:papost@cc.uoa.gr, mtsampa@cc.uoa.gr} and P. S. Apostolopoulos$^{*}$}
\maketitle

\begin{abstract}
We present two counter examples to the paper by Carot et al. in Gen. Rel.
Grav. (1997). {\bf 29 }1223{\bf \ } and show that the results obtained are
correct but not general.
\end{abstract}


Suppose ($M_{1},h_{1}(x^{A})$) and ($M_{2},h_{2}(x^{\alpha })$) are a pair
of Riemannian manifolds with co-ordinate functions $x^{A}$ ($A,B,...=1,2$)
and $x^{\alpha }$ ($\alpha ,\beta ,...=3,4$) respectively. Let $\Phi (x^{C})$
be a real valued function on $M_{1}$ and $M=M_{1}\times M_{2}$ be the
product manifold (type B warped space-time) with metric \cite{Carot-da Costa}%
:

\begin{equation}
ds^2=h_{AB}(x^C)dx^Adx^B+\Phi ^2(x^C)h_{\alpha \beta }(x^\gamma )dx^\alpha
dx^\beta .  \label{sx1}
\end{equation}
A vector field ${\bf X}$ in $M$ can be decomposed uniquely in ''horizontal''
and ''vertical'' components as follows:

\begin{equation}
X^a=X_1^A(x^b)\delta _A^a+X_2^\alpha (x^b)\delta _\alpha ^a  \label{sx2}
\end{equation}
where $a,b,...=1,2,3,4$.

In a recent paper \cite{J Carot-Nunez-Percoco} (referred from now on as CNP)
Carot et al. have considered the problem of determining all Ricci
collineations (RCs)\ of type B warped space-times and have come to the
following conclusion:

{\em The horizontal component }${\bf X}_2${\em \ of a {\bf proper} RC {\bf X}
in a warped type B space-time is either a Homothetic Vector Field (HVF) of }$%
(M_2,h_2)$ {\em and {\bf X} is given by: }

\begin{equation}
{\bf X}=X_1^A(x^a)\partial _A+C^I(x^B){\bf \zeta }_I  \label{sx3}
\end{equation}
$C^I${\em \ and }$X_1^A${\em \ being functions of their arguments to be
determined from the relations}

\begin{equation}
R_{AB,D}X_1^D+R_{AD}X_{1,B}^D+R_{DB}X_{1,A}^D=0  \label{sx4}
\end{equation}

\begin{equation}
R_{AD}X_{1,\alpha }^D+Fh_{\alpha \beta }X_{2,A}^\beta =0  \label{sx5}
\end{equation}
{\em and \{}${\bf \zeta }_I${\em \}\ (with }$I\leq 4${\em ) form a basis of
the homothetic algebra of }$(M_2,h_2{\em )}$ {\em or }

${\bf X}_2$ {\em it is a proper Special Conformal Killing Vector (SCKV) of }$%
(M_2,h_2),${\em \ this being possible only when }($M_2,h_2$){\em \ is flat,
in which case }$X${\em \ is given by:}

\begin{equation}
{\bf X}=\left[ P^{A}(x^{B})x+Q^{A}(x^{B})\right] \partial _{A}+\left[ \frac{%
A_{0}}{2}(x^{2}-y^{2})-D_{0}y+L(x^{B})\right] \partial _{x}+\left(
A_{0}xy+D_{0}x\right) \partial _{y}  \label{sx6}
\end{equation}
{\em where the functions specifying }${\bf X}${\em \ must satisfy (\ref{sx4}%
), (\ref{sx5}) and }$x,y${\em \ are co-ordinate functions in }$(M_{2},h_{2}).
$

Unfortunately the conclusion above is not the solution to the problem
considered, in the sense that it does not give all (smooth) RCs of a type B
warped space-time. We show this by giving the following two counter examples.

\underline{Counter example 1}

Consider a globally $\{2+2\}$ decomposable space-time i.e. $\Phi (x^C)=1$.
It is always possible to choose co-ordinates $x^a=\left\{ t,x,y,z\right\} $
in which the metric is written:

\begin{equation}
ds^2=f^2(x^C)(-dt^2+dx^2)+f^{\prime 2}(x^\gamma )(dy^2+dz^2)  \label{sx7}
\end{equation}
where the functions $f,f^{\prime }$ are smooth functions of their arguments.

Consider the vector ${\bf X}=\partial _t+\partial _y$ and define the
functions $f,f^{\prime }$ to be:

\begin{eqnarray}
f &=&e^{\frac{a_1t^2}2+b_1t+c_1}  \nonumber \\
&&  \label{sx8} \\
f^{\prime } &=&e^{\frac{a_2y^2}2+b_2y+c_2}  \nonumber
\end{eqnarray}
It is easy to show that for this choice of $f,f^{\prime }$ the curvature
scalars $R_1=2a_1e^{-a_1t^2-2b_1t-2c_1}$, $R_2=-2a_2e^{-a_2y^2-2b_2y-2c_2}$
do not vanish and furthermore ${\cal L}_{{\bf X}}R_{ab}=0$, ${\cal L}_{{\bf X%
}}g_{ab}\neq 2\psi ({\bf X})g_{ab}$ and ${\cal L}_{{\bf X}}R_{bcd}^a\neq 0$
so that ${\bf X}$ is a proper (smooth) RC of the 2+2 metric $g_{ab}$. Taking
the Lie derivative of the 2-metrics with respect to the projections ${\bf X}%
_1=\partial _t,$ ${\bf X}_2=\partial _y$ of ${\bf X}$ in the 2-spaces we
find:

\begin{eqnarray*}
{\cal L}_{{\bf X}_1}h_{AB} &=&2(a_1t+b_1)h_{AB}=2\psi _1({\bf X}_1)h_{AB} \\
&& \\
{\cal L}_{{\bf X}_2}h_{\alpha \beta } &=&2(a_2y+b_2)h_{\alpha \beta }=2\psi
_2({\bf X}_2)h_{\alpha \beta }
\end{eqnarray*}
which show that ${\bf X}_1,{\bf X}_2$ are{\it \ }{\em proper}{\it \ }CKVs of
the 2-spaces $(M_1,h_1)$ and $(M_2,h_2)$ respectively contrary to the
conclusions of CNP. Furthermore it is easy to show that ${\bf X}_1,{\bf X}_2$
are proper Ricci collineations of the corresponding 2-metrics. [It is well
known \cite{Hall-Roy-Vaz} that in a 2+2 decomposable space-time the sum $%
{\bf X}={\bf X}_1+{\bf X}_2$ of Ricci collineations ${\bf X}_1,{\bf X}_2$ of
the 2-spaces $(M_1,h_1)$ and $(M_2,h_2)$ respectively, defines a Ricci
collineation for the whole space-time.]

\underline{Counter example 2}

In the 2+2 decomposable space-time given above (and the same functions $%
f,f^{\prime })$ consider the vector field ${\bf X}=z\partial _{t}+\partial
_{y}+t\partial _{z}$. It is easy to show that ${\cal L}_{{\bf X}}R_{ab}=0$
provided $a_{1}=-a_{2}$. Furthermore ${\cal L}_{{\bf X}}g_{ab}\neq 2\psi (%
{\bf X})g_{ab}$ and ${\cal L}_{{\bf X}}R_{bcd}^{a}\neq 0$. The projection $%
{\bf X}_{1}=z\partial _{t}$ is a {\em proper} CKV for the Lorentzian 2-space
with conformal factor $\psi ({\bf X}_{1})=z(b_{1}-a_{2}t)$ and the
projection ${\bf X}_{2}=\partial _{y}+t\partial _{z}$ is a {\em proper} CKV
of the Euclidean 2-space with conformal factor $\psi ({\bf X}%
_{2})=a_{2}y+b_{2}$ where both 2-spaces are not flat. Due to the dependence
of ${\bf X}_{1}$ from $x^{4}$ and ${\bf X}_{2}$ from $x^{1}$, this
counterexample also shows, that the Ricci collineations ${\bf X}_{1},{\bf X}%
_{2}$ are ${\em not}$ ${\em invariant}$ under the isometries linking the
timelike (spacelike) submanifolds of the decomposition \cite{Hall-Roy-Vaz}.

As we have remarked the results obtained in CNP are correct but they are not
general. This is due to the basic relation (17) of CNP which {\em is not true%
} for a 2-dimensional space.

Indeed it can be proved that if ${\bf X}$ is a proper CKV of a $n-$%
dimensional space with metric $g_{ab}$ satisfying ${\cal L}_{{\bf X}%
}g_{ab}=2\psi ({\bf X})g_{ab}$ then the following identity holds:

\begin{equation}
{\cal L}_{{\bf X}}R_{ab}=-(n-2)\psi _{;ab}-\left( \psi _{;cd}g^{cd}\right)
g_{ab}  \label{sx8a}
\end{equation}
where $R_{ab}$ is the Ricci tensor and a semicolon denotes covariant
differentiation associated with the $n-$dimensional metric $g_{ab}$.

Hence for a proper CKV ${\bf X}_2$ of a 2D metric $h_{\alpha \beta }$ i.e. $%
{\cal L}_{{\bf X}_2}h_{\alpha \beta }=2\Psi ({\bf X}_2)h_{\alpha \beta }$
equation (\ref{sx8a}) implies:

\begin{equation}
{\cal L}_{{\bf X}_2}R_{2\alpha \beta }=-\left( \Psi _{|\rho \sigma }h^{\rho
\sigma }\right) h_{\alpha \beta }  \label{sx9}
\end{equation}
where a stroke denotes covariant differentiation w.r.t. 2-metric $h_{\alpha
\beta }$.

Thus, equation (21) of CNP does not follow from (17) but can be seen rather 
{\em as an extra assumption} which is consistent with equation (\ref{sx9}).

To show this consider equation (\ref{sx9}). Since for every 2-space $%
R_{2\alpha \beta }=\frac{R_2}2h_{\alpha \beta }$ equation (\ref{sx9})
implies:

\begin{equation}
{\bf X}_{2}\left( R_{2}\right) +2\Psi R_{2}=-2\left( \Psi _{|\rho \sigma
}h^{\rho \sigma }\right) .  \label{sx10}
\end{equation}
Contraction of equation (21) of CNP together with (\ref{sx10}) gives ${\bf X}%
_{2}\left( R_{2}\right) +2\Psi R_{2}=0$. Thus either $R_{2}=0$ or ${\bf X}%
_{2}\left( R_{2}\right) =-2\Psi R_{2}$ where the later equation gives $\Psi
_{,\alpha }=0$ i.e. ${\bf X}_{2}$ is a HVF for the metric $h_{\alpha \beta }$%
. 

We conclude that the problem of determining the (proper and smooth) RCs of
type B warped space-times is still open although a serious step towards its
solution has been done in CNP.

\end{document}